# ARTICLE

# On the switching mechanism and optimisation of ion irradiation enabled 2D MoS2 memristors

Samuel Aldana,*[a] Jakub Jadwiszczak [a] and Hongzhou Zhang [a]

Memristors are prominent passive circuit elements with promising futures for energy-efficient in-memory processing and revolutionary neuromorphic computation. State-of-the-art memristors based on two-dimensional (2D) materials exhibit enhanced tunability, scalability and electrical reliability. However, the fundamental of the switching is yet to be clarified before they can meet industrial standards in terms of endurance, variability, resistance ratio, and scalability. This new physical simulator based on the kinetic Monte Carlo (kMC) algorithm reproduces the defect migration process in 2D materials and sheds light on the operation of 2D memristors. The present work employs the simulator to study a two-dimensional 2H-MoS$_2$ planar resistive switching (RS) device with an asymmetric defect concentration introduced by ion irradiation. The simulations unveil the non-filamentary RS process and propose routes to optimize the device's performance. For instance, the resistance ratio can be increased by 53% by controlling the concentration and distribution of defects, while the variability can be reduced by 55% by increasing 5-fold the device size from 10 to 50 nm. Our simulator also explains the trade-offs between the resistance ratio and variability, resistance ratio and scalability, and variability and scalability. Overall, the simulator may enable an understanding and optimization of devices to expedite cutting-edge applications.

## 1. Introduction

The development of Information and Communication Technology (ICT), such as 5G circuits and the internet of things, demands breakthroughs in non-volatile memory[1] since the state-of-the-art flash technology is hitting its physical limits in terms of power consumption and device scalability.[2] Among the emerging memory technologies, memristors have attracted much research interest and become the herald of next-generation computational architecture, revolutionizing ICT.[3-6] The potential of memristors is rooted in their superior performance (e.g., ultrafast switching, low power consumption, data retention and endurance), intrinsically high scalability, stackability, compatibility with Complementary Metal Oxide Semiconductor (CMOS) technology and flexibility for wearable applications.[1,2,7] However, device variability, including cell-to-cell variability and cycle-to-cycle variability, hinders the industrial deployment of memristor technology. The cell-to-cell variability is the inhomogeneity between devices via the same fabrication process and cycle-to-cycle variability is related to the operation of individual devices,[1,7] which emerges from the stochastic processes during the resistive switching (RS). For example, in filamentary RS devices the location and morphology of conducting filaments may vary between cycles and cells.[8,9] Device variability is inevitable in RS devices which relies on material defects[10-13], while the forming process exacerbates the problem by varying defect concentration and distribution[14]. Although the cycle-to-cycle variability can be insignificant in some applications exemplified by neuromorphic imaging recognition,[1] it imposes major limitations on memristor applications, rendering a range of high-end applications impractical (e.g., multilevel information processing and long-term storage[7]). For example, device variability causes signal degradation in crossbar arrays[15] and hampers projections of device lifetime, demanding excessive budget in testing.[16] Verification and iterative approaches may mitigate the resistance variability in multilevel information[17] and radiofrequency applications respectively.[18] However, a consensus on a figure of merit for the variability issue is yet to be achieved and a lack of comparable statistics on device variability remains a main obstacle to the technology.[7]

Recently, memristive behaviour has been observed in a range of two-dimensional (2D) materials.[19-25] This may expedite the industry deployment of the memristor technology since 2D memristors exbibit superior tunability, scalability and electrical reliability.[4,19,26-34] These characteristics arise from the ultrathin nature and unique mechanical, electric and optoelectronic properties of 2D materials.[26,27] The 2D memristor landscape shows diverse device architectures and switching mechanisms. For example, the switching of 2D vertical memristors[20,35-38] depends on the formation and rupture of conductive filaments. 2D planar memristors may rely on phase transition,[39,40] charge trapping/de-trapping,[41,42] electron tunnelling modulated by polarization,[43] electrochemical processes[44] and Schottky barrier modulation.[29] Defect migration plays a crucial role in these processes.[32,45-47] Planar 2D memristors show great promise in

[a.] Centre for Research on Adaptive Nanostructures and Nanodevices (CRANN), Advanced Materials and Bioengineering Research (AMBER) Research Centers, School of Physics, Trinity College Dublin, Dublin, D02 PN40, Ireland. E-mail: aldanads@tcd.ie

Electronic Supplementary Information (ESI) available: See DOI: 10.1039/x0xx00000x

**ARTICLE**  **Journal Name**

the future of energy-efficient neuromorphic computations. The planar architecture allows effective gate tuning and multi-terminal operation, enhancing device controllability and enabling complex neural functionalities. [32] Their low intrinsic capacitance supports fast switching and low power consumption. We note the planar 2D memristor is still in its infancy. The prototype devices operate at much higher voltages (∼10s V) than the state-of-the-art vertical 3D memristors (∼1-2V). It is imperative to improve the device performance. Compared with their 3D counterparts, the device variability of 2D memristors has rarely been explored, while it is of utmost importance to gain in-depth knowledge of the switching process and mitigate the variability issues in 2D memristors.

Physical simulators are indispensable for understanding the resistive switching process and they can greatly facilitate investigations on device viability. [7, 48, 49] Ab initio calculations can accurately relate the resistive switching to the defect creation, [50] the electronic structure and transport properties in nanostructures. [51] However, they are limited to relatively small volumes (a few nm as maximum) and short times (shorter than ns) and can hardly reproduce RS processes at the device level. [48] Continuous models are apt to describe the average behaviours of individual devices or even circuits. However, they overlook the microscopic characteristics of the system, such as particle migration[52-55], and are hence not suitable to investigate the stochasticity that emerges from the evolution of microscopic configuration of the system during the switching process. The kinetic Monte Carlo (kMC) algorithm is an established technique to study the microscopic evolution and its impacts on device performance. [11, 12, 56-59] For example, the kMC algorithm can reproduce particle diffusion and the formation and rupture of percolation paths involving several RS cycles of memristors. [11, 56-58] Although the kMC algorithm and continuous models have been employed to explore defect accumulation and electrical conduction in 2D materials, [29, 38] the simulation of 2D memristors at the device level is scarce and the variability of 2D memristors has been rarely explored. In this work, we build a new physical kMC simulator for the defect migration process in 2D materials and shed light on the operation of 2D memristors. We investigate the effects of initial vacancy distribution, the scaling limits and the factors that regulate device endurance and variability. We use experimental data from the $MoS_2$ memristor enabled by site-specific defects to collate and verify the model. [19] The simulator helps understand the physics of resistive switching in 2D materials, offering practical guidance to optimize device performance.

## 2. Experimental results and simulation approach

**Figure 1a** shows a device schematic. The device exhibits a planar metal-semiconductor-metal structure and the semiconductor channel is a 2D 2H-$MoS_2$ with an asymmetric defect concentration introduced by focused helium ion irradiation (see more details about the device fabrication in Supplementary Information, section 1). We note that the irradiation-induced defects enable the resistive switching, while devices of pristine $MoS_2$ do not switch. [19] The defect region, referred to as the fissure, bisects the channel with an asymmetric concentration along the horizontal direction, i.e., across the electrodes. For example, the right tail of the peak (tens of nanometres) in Figure 1a is much wider than the left (< 1 nm).

In our simulation, we have assessed three initial defect distributions with different asymmetries (see Supplementary Information, section 2), i.e., a skewed Gaussian distribution (**Figure S1a**), a triangle function (Figure S1b) and a step function (Figure S1c). The skewed Gaussian distribution emulates the asymmetric distribution of defects observed experimentally[19], and the triangle and step functions of different asymmetries are to explore possible routes for device optimization. The source electrode as convention is grounded and placed at the left side of the asymmetric peak. The polarity of applied voltage regulates the direction of the fields with respect to the asymmetric defect distribution, i.e., a positive voltage ($V_d > 0$) indicates the electric field points from the abrupt edge of the fissure to the long tail side.

Figure 1b shows 14 experimental quasi-static pinched hysteresis loops by selecting every 15th loop from 1200 consecutive cycles measured at a triangular voltage ramp between ±35 V with a rate of $6\,V\,s^{-1}$. The device shows bipolar operation, and the SET (RESET) process occurs at a positive (negative) voltage. The current level increases with the cycling. Figure 1c shows the resistance ratio over cycle, which shows that cycling the device leads to an exponential decay (shown as a dashed line) in resistance ratio (21% after 1000 cycles). These degradation processes will be further discussed alongside the simulation work.

To investigate the device switching process, we implemented the kinetic Monte Carlo (kMC) algorithm[60] using MATLAB in a simulation domain of $50 \times 50$ nm. The vacancy distribution within the fissure evolves with the external electric field applied via the electrodes,[19] while no defects escape the simulation domain during the switching. Since the fissure region dominates the device resistance, the size of the simulation domain is sufficient to include the main physical processes involved in the resistive switching. The defects are doubly charged sulfur vacancies because the helium-ion irradiation preferentially removes sulfur atoms from the $MoS_2$ lattice.[33] The generation of new defects under the applied electric field is negligible due to the high activation energy ($5.85$ eV[61] for vacancy and $> 5$ eV for antisite defect[62]). The sulfur vacancy migrates via the exchange of the vacancy position with one of the adjacent sulfur atoms,[29, 63] as shown in the Figure 1a. In the quasi-static switching, the vacancy migration events occur at a much longer time scale ($\sim$ s) than the lattice vibration ($10^{-13}$ s), so the system is in thermodynamic equilibrium for any vacancy distributions.[60] Furthermore, the field-driven migration renders the reverse migration negligible, validating the kMC approach. In our simulation, we combine the electric response with the thermal effect since the lattice temperature modulates the transition rate, which is given by Maxwell-Boltzmann statistics and Transition State Theory (TST), i.e. $\Gamma = \nu \cdot \exp(-E_A/K_B T)$, where $\nu = 7 \times 10^{13}$ s$^{-1}$ is the vibration constant of the particle, $K_B$ the Boltzmann constant, $T$ the temperature and $E_A$ the activation energy of the migration. The activation energy is modulated by the local electric field as follows: $E_A = E_A^0 - \boldsymbol{b} \cdot \boldsymbol{F}(x,y)$, [12] where $\boldsymbol{F}(x,y)$ is the electric field, $\boldsymbol{b}$ the polarization factor and $E_A^0 = 2.297$ eV the activation energy for migration in the zero-field condition. [4] The kMC algorithm weighs all the possible migrations and chooses the evolvement path. It should be noted that the larger the transition rate, the smaller the time step $t = -\ln(rand)/\sum \Gamma$, where $rand$ stands for a random number between 0 and 1 and $\sum \Gamma$ is the summation of the transition rates for all possible migrations. For each defect distribution during the switching, the local vacancy concentration $\rho_d(\vec{r})$ is averaged over $6 \times 6$ grid points (3.2 nm²), which determines the local resistance $R(\vec{r})$ via the empirical relationship $R(\vec{r}) \propto \rho_d^n(\vec{r})$, [33] ($n$ is a parameter extracted from the experimental results, see Supplementary Information, section 4). The electric field screening is assumed to be a linear function of the local resistance since the dielectric constant in MoS2 strongly depends on the distribution and number of sulfur vacancies. [64] For a given applied voltage, the local electric field decreases with the increase in the defect density, indicating the defect migration is a self-limiting process. Further details about the simulation can be found in the Supplementary Information, section 3.

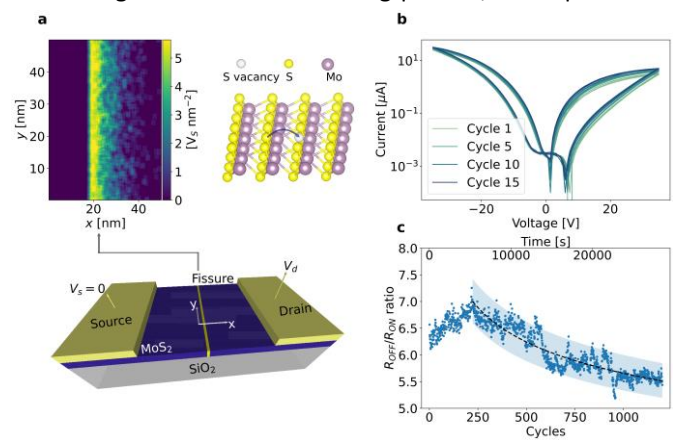

**Figure 1:** Defect enabled **memristive device.** a) Schematic device structure shows the fissure of defects in the $MoS_2$ channel. The $50 \times 50$ nm simulation domain represents the defect profile of a skewed Gaussian distribution, while the channel is on the micrometre scale. The arrows in the lattice structure indicate the possible routes of vacancy migration, i.e., a S atom (yellow) exchanges its position with a nearest-neighbour vacancy site (grey). b) 14 representative experimental I-V hysteresis loops sampled from continuous 1200 cycles. c) The evolution of the resistance ratio (calculated at -10 V) over cycles (time in the top *x*-axis). The dashed line is an exponential fitting, and the shaded indicates the 95% prediction interval.

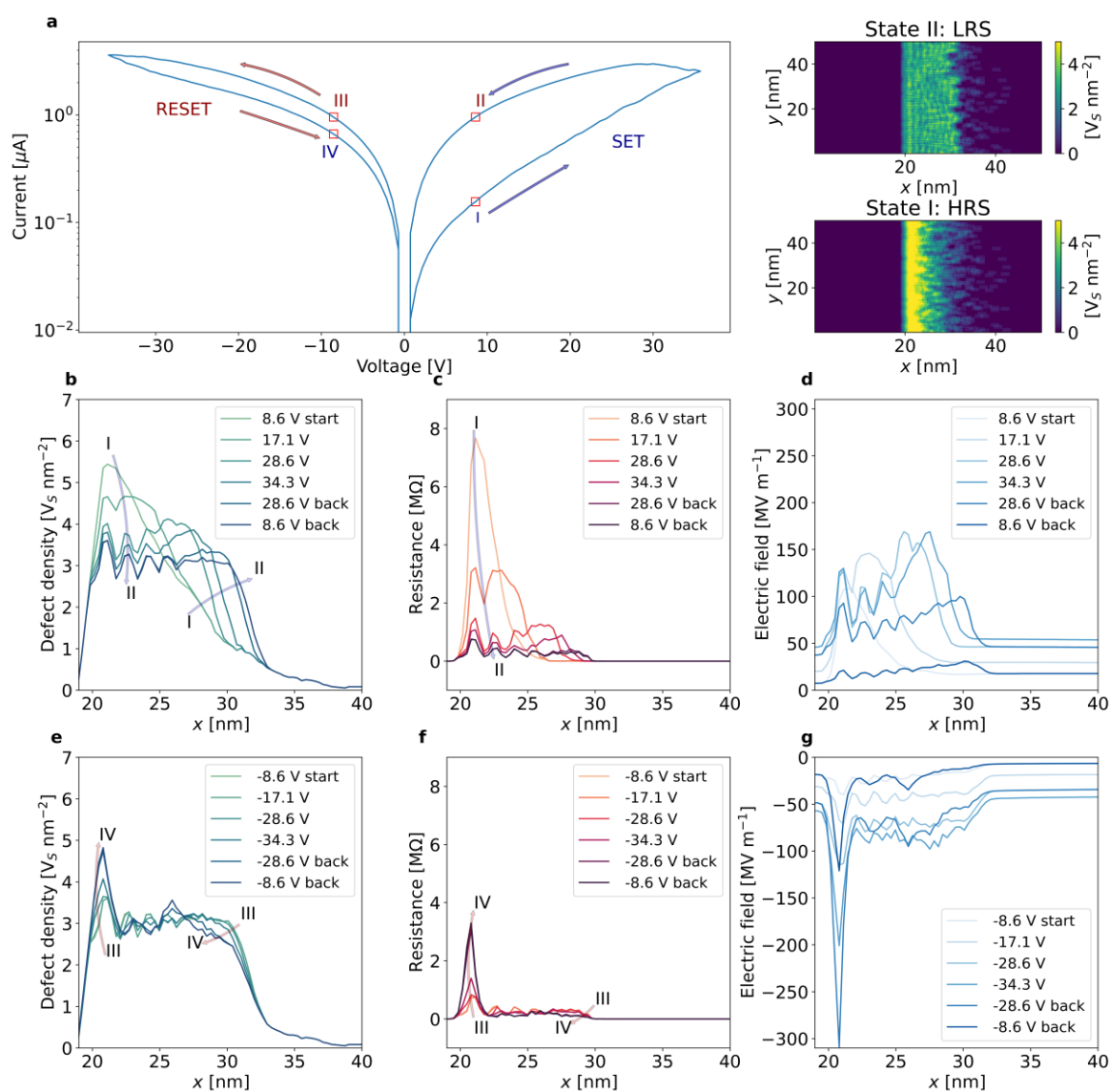

**Figure 2: Switching mechanism.** The initial defect profile of the device exhibits a skewed Gaussian distribution with a peak density of 5.64 $V_s$ nm$^{-2}$ and a width of 8 nm. The voltage ramp starts from positive polarity with a rate of 0.71 V s$^{-1}$ between 35 V and −35 V. a) a typical simulated I-V pinched hysteresis loop. Numerals on the loop mark four representative states of the switching process. The SET process takes place from I to II, and for the RESET from III to IV. The colour maps shown in the right panel correspond to the microscopic configuration of defects in the LRS (top) and HRS (bottom). b), c) and d) are the defect density, local resistance and electric field profiles along the x-axis during the SET, respectively. e), f) and g) are the correspondent profiles during the RESET processes.

## 3. Simulation results and discussions

**Figure 2a** shows a typical simulated pinched hysteresis loop from a defect distribution of skewed Gaussian profile under a voltage ramp rate of 0.71 V s$^{-1}$ between 35 V and -35 V. Prior to the switching, the fissure region is 8 nm wide with a peak density $\rho = 5.64$ V$_s$ nm$^{-2}$ (see the probability distributions in Supplementary Information, section 2). The I-V sweep follows the directions indicated by the arrows. The loop shows a SET process with a positive voltage and a RESET process with a negative voltage, suggesting the same bipolar switching behaviour observed experimentally. The simulation corroborates that the operation of the device does not need a forming process, facilitating circuit simplicity. [65, 66] The switching is progressive, in contrast to an abrupt resistance change, suggesting the absence of filamentary conduction. This explains the observed low power consumption. [7] The resistance ratio ($r = R_{off}/R_{on}$) is 1.44 calculated at -4 V during the RESET process and the maximum current level is 3 μA. It is interesting to note the simulated loop reproduces the asymmetry found experimentally between the SET and RESET processes (see Figure 1b).

The switching is due to the reconfiguration of the defect distribution within the fissure by the electric field. The two colour-maps attached to Figure 2a show the microscopic distributions of defects in the High Resistance State (HRS) and the Low Resistance State (LRS), corresponding to the state labelled by I) and II) on the loop respectively (more microscopic details about the states I-IV can be found in Supplementary

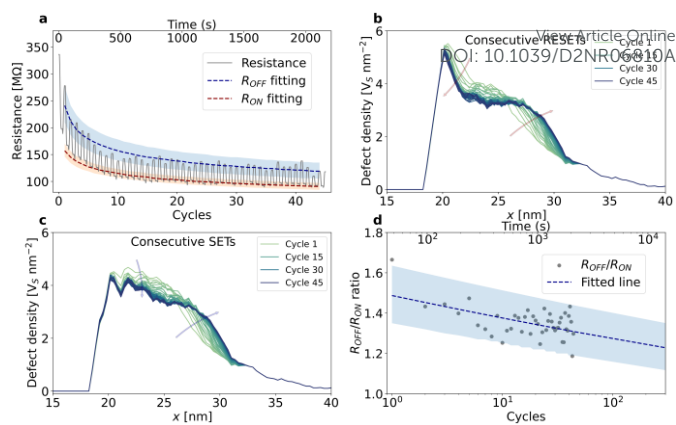

**Figure 3: Device fatigue.** The initial defect profile of the device exhibits a skewed Gaussian distribution with a peak density of 5.64 V$_s$ nm$^{-2}$ and a width of 9 nm. The voltage ramps from a positive polarity with a rate of 2.1 V s$^{-1}$ between 25.2 V and −25.2 V for 45 cycles (2161 s). a) Resistance evolution over time with two exponential fittings for the LRS (in red) and the HRS (in blue) of the cycles. The shaded regions correspond to the 95% prediction interval of the fitting. b) and c) correspond to the density profile along the x-axis (averaged over the y-axis) for successive RESETs and SETs respectively. d) resistance ratio projection based on the simulation.

Information, section 5). The defects accumulate within a small region in the HRS, leading to a much higher density than the LRS. Figure 2b details the defect evolution during the SET process. The defect profiles are extracted at a series of sequential voltages from state I to II (see Figure 2a) and the lightness of the curves reduces with increasing time. The initial vacancy concentration exhibits a skewed Gaussian distribution (the most intense red curve), mimicking the defect profile by the ion irradiation. The external field of positive polarity gradually drives the defects away from the fissure region, lowering the peak and leading to a plateau of the distribution across a 10-nm wide region. The local resistance varies with the defect profile (see Figure 2c). The spatial-varying resistance regulates the electric potential distribution in the channel when an external voltage is applied (see Figure 2d). The larger the resistance of a region, the larger the electric field, and the more significant the vacancy drifting. Therefore, the self-adaptive electric field reduces the vacancy concentration and hence the resistance. From State I to II, the peak resistance reduces by an order of magnitude, leading to an overall reduction in the channel resistance and hence the SET process.

Figure 2e reveals the evolution of the defect profile during the RESET process from state III to IV (see Figure 2a). The defects move towards and accumulate at the abrupt edge of the fissure ($x = 22$ nm in Figure 2e), which increases the local resistance (see Figure 2f) and hence the local electric field (see Figure 2g). On the long tail of the peak (i.e., $x > 22$ nm), the field pushes defects from a wide region (22 nm $< x <$ 30 nm) towards the peak, while on the left side of the peak the field drops drastically within a 1 nm region. This asymmetric distribution of the field limits the escape of the defects from the fissure into the left side of the channel and causes the defect accumulation at the peak, recovering the initial defect configuration. The drift of defects to the left side of the fissure may become important if the field is sufficiently high where the RESET process will fail, leading to device failure (see **Figure S4**). The simulator can predict the

Journal Name

ARTICLE

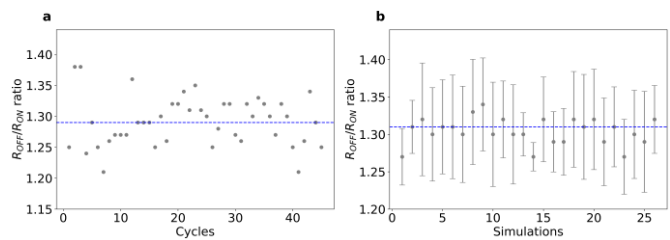

**Figure 4: Device variability.** The initial defect profile of the device exhibits a skewed Gaussian distribution with a peak density of 5.64 $V_s$ nm$^{-2}$ and a width of 9 nm. The voltage ramps from a negative polarity with a rate of 2.1 V s$^{-1}$ between 25.2 V and -25.2 V. a) shows simulated cycle-to-cycle variability of the resistance ratio with an average of 1.29 (the dashed line). b) corresponds to the resistance ratio variability of 26 independent 15-cycle simulations initiated with the same parameters but varying microscopic defect configurations. For each simulation, the resistance ratio is the average over the 15 cycles and the error bar is the standard deviation of the cycle-to-cycle variability.

operational range of the voltage for the device. For example, a device with a peak density of 5.64 $V_s$ nm$^{-2}$ and a width of 12 nm can be stressed with -30 V for 17 s or -40 V for 2 s before the device fails (see Figure S4a).

The simulator allows us to investigate the device performance. **Figure 3a** shows the temporal evolution of the device resistance to a triangular voltage ramp (2.1 V s$^{-1}$) between 25.2 and −25.2 V. The initial defect profile has a skewed Gaussian distribution with a peak density of 5.64 $V_s$ nm$^{-2}$ and a width of 9 nm. The voltage ramp starts from a positive polarity $+Vd$. The device switches over 45 cycles (2161 s). Both the $R_{on}$ and $R_{off}$ drop 57% after the first ten cycles and reaches a steady state where $R_{on}$ and $R_{off}$ drop slower (17% in 35 cycles). The resistance drop occurs due to a progressive reduction in the peak density of the defect profile, which dominates the device resistance. This is evident in Figure 3b and 3c, which show the evolution of the density profile with consecutive RESET and SET processes. Here, we can see that cycling gradually relocates the defects into the originally low-density (right tails) regions (below 3.5 $V_s$ nm$^{-2}$) at the expense of the peak density. This reduces the defects population of the peak region from 31% in cycle 1 to 22% in cycle 45, as can be seen in Figure S5a. The defect accumulation in the tail region (> 25 nm onward in the *x*-axis) stems from the low electric field in the region, which is 16% of the electric field found in the peak (see Figure S5b). The low field modulates the migration barrier by 1% (see Figure S5c), leading to negligible field-driven migration in the tail region. The further the defects migrate away from the peak into the tail during the SET process, the harder for them to move back to the peak during the RESET process. Consequently, the defect distribution flattens across the fissure during the cycling leading to a gradual reduction in the device resistance. This flattening of the defect profile progressively reduces the difference between the OFF and ON states, reducing the ON/OFF ratio and eventually leading to device failure (see also the resistive switching loops in Figure S6). As observed experimentally, the power consumption increases (see the rise from 20 μW to more than 50 μW in 45 cycles in Figure S7) with the resistance reduction. Nevertheless, the simulation qualitatively explains the fatigue behaviour (see Figure 1c) and allows quantitative prediction. For devices with long cycling tolerance, we can fit the behaviours of $R_{on}$ and $R_{off}$ (red and blue dashed lines in Figure 3a respectively) and project the evolvement of ON/OFF ratio for longer times (Figure 3d).

The most significant advantage of the simulator is its capability of exploring device variability since the kMC algorithm is apt to investigate stochastic processes. **Figure 4a** shows the cycle-to-cycle variability of the resistance ratio in one 45-cycle simulation, starting with a skewed Gaussian distribution of vacancies, negative polarity, with a peak density $\rho_1 = 5.64$ $V_s$ nm$^{-2}$ and a width of 9 nm. The value of resistance ratio distributes uniformly in the range of 1.2 to 1.4 with the mean of 1.29. We define the cycle-to-cycle variability of two consecutive cycles as $\Delta r = |r_i - r_{i+1}|$. The standard deviation of the cycle-to-cycle variability is 0.03. In Figure 4b, we investigate the cell-to-cell variability by running 26 independent simulations for a given macroscopic distribution (see section 3 in Supplementary Information for more details about running different simulations). Each simulation starts with a unique initial microscopic vacancy configuration in the lattice, while the macroscopic defect distribution remains the same. This scenario mimics the device fabrication process, where the sputtering process and the creation of vacancies are stochastic at the nanometre scale, [19, 33] introducing cell-to-cell variability. We note the device fabrication is limited by many other

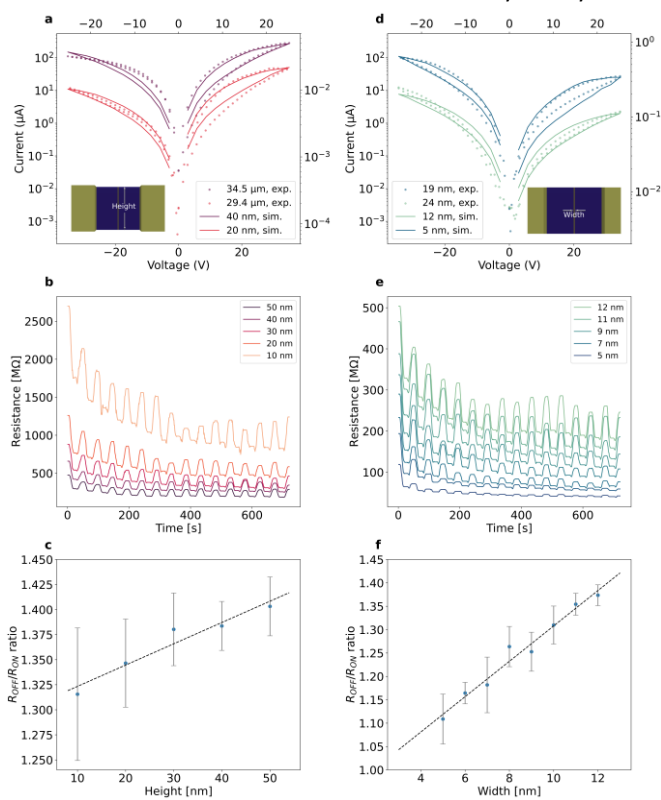

**Figure 5: Device scalability.** a), b) and c) the effects of the device height on the switching d), e) and f) the effects of the fissure width on the switching. In a) and d), the experimental (simulated) data are plotted against bottom (top) and left (right) axes. For the simulation, the initial defect profile of the device exhibits a skewed Gaussian distribution with a peak density of 5.64 $V_s$ nm$^{-2}$. The voltage ramps from a positive polarity with a rate of 2.1 V s$^{-1}$ between 25.2 V and -25.2 V for 15 cycles. The resistance ratio is averaged over the 15 cycles and the error is the standard deviation of the cycle-to-cycle variability.

**ARTICLE**  **Journal Name**

parameters (e.g., ion beam stability, focusing, sample cleanliness, etc.) and the cell-to-cell variability here represents the upper limit of the ideal situation. For each such simulation, we average the resistance over 15 cycles and use the standard deviation of the cycle-to-cycle variability as the error. Here, we can observe a uniform distribution of resistance ratios around the mean of 1.31 with a standard deviation of 0.02.

We investigate the potential of device scalability by evaluating the effects of the device dimensions (i.e., height and width) on the device resistance and the resistance ratio. For the simulations shown in **Figure 5**, the initial defect profile has a skewed Gaussian distribution with a peak density of 5.64 $V_s$ nm$^{-2}$. The device is under a voltage ramp rate of 2.1 V s$^{-1}$ between 25.2 and -25.2 V, which starts from the positive polarity. Figure 5a displays experimental and simulated pinched I-V hysteresis loops with varying device height, while Figure 5b shows the simulated temporal evolution of resistance. Both the experiment and simulation show the decrease in device height results in an overall resistance increase. This is due to the reduced conducting channels along the vertical direction. This corroborates the non-filamentary conduction. Figure 5c demonstrates the limit of scaling down the vertical dimension. Both the resistance ratio and the device variability deteriorate as the device height reduces. A trade-off needs to be identified between the device conductance and the resistance ratio (variability), indicating a limit on the scalability of the *y*-dimension. Scaling along the horizontal dimension is regulated by the fissure width and the range of defect drift. Figure 5d shows typical experimental and simulated I-V hysteresis loops from devices with varying fissure width. The temporal resistance evolution is simulated in Figure 5e. The device resistance decreases with decreasing the fissure width, but the resistance ratio and device variability deteriorate as the fissure width is scaled down (Figure 5f). Our simulations reveal the limits of device scaling.

The asymmetric nature of the initial defect distribution is crucial to the resistive switching. This indicates that device optimization may be possible by tuning the initial defect distributions. In **Figure 6** we show the resistance ratio for an initial defect distribution with a triangle shape (Figure 6a) and a step function shape (Figure 6b) with varying peak densities and fissure region widths. The skewed Gaussian distribution case is in **Figure S8**. In the triangle case (Figure 6a), when the peak density is higher than 3.95 $V_s$ nm$^{-2}$, the resistance ratio appears to exhibit a maximum when the fissure width is varied from 4 nm to 32 nm. The value of the maximum ratio decreases and occurs at a larger fissure width when the peak density decreases. Devices with the step function distribution (Figure 6b) exhibits a similar maximum ratio. In contrast to the triangle case, the maximum appears in all the peak densities simulated and on the right side of the peak the resistance ratio falls more rapidly with increase in the fissure width compared with the triangle distribution. However, the triangle distribution enables a higher resistance ratio at a narrower fissure than both the Gaussian and step function cases, so it may offer a better option for device scaling and further performance optimization. We also note that in all the distributions simulated, the resistance ratio and variability can be enhanced by increasing the initial peak density. This is because a larger defect population enables more significant differences between the HRS and LRS.

Finally, we note that although the voltage is high in this MoS$_2$-based device, the current is low, so the device energy consumption is reasonable. Besides, as the voltage drops mainly in the fissure region, we cannot address the scaling of the switching voltage by reducing the length between the drain and the source. In this sense, the minimum electric field needed to move the vacancies determine the voltage scale. Hence, it might be possible to lower the operating voltage and further reduce the power consumption by increasing the density peaks (see in Figure S4 how the electric field has a stronger influence in higher densities), selecting materials with suitable activation energies[63] or by defect engineering (defects migrate easier through grain boundaries[4]).

## Conclusions

We have developed a kMC simulator for 2D planar memristors based on defect migration using the case of a MoS$_2$-based device enabled by a Helium Ion Beam Microscope. The simulator reproduces the asymmetric resistive switching cycle with a performance close to observed experimentally. Besides, this approach is helpful for insights into the switching mechanism, in addition to study the device variability, the device endurance and the resistance ratio. We studied the device degradation with cycling, which produces defect relocations into low-density regions, causing a drop in the resistance ratio and reducing the switching window. Some trade-offs are proposed to tailor the device features by controlling the device size, the number of defects introduced in the channel and their distribution. For example, the device can be miniaturized at the expense of reducing the resistance ratio, increasing the variability and the resistance, or higher peak densities can be used to increase the resistance ratio and variability. We note that to reduce the switching voltage of a device based on defect migration, we must enhance the migration of defects by employing defect engineering, using higher peak densities or other Transition Metal Dichalcogenides with lower activation energy for defect migration. Besides, the

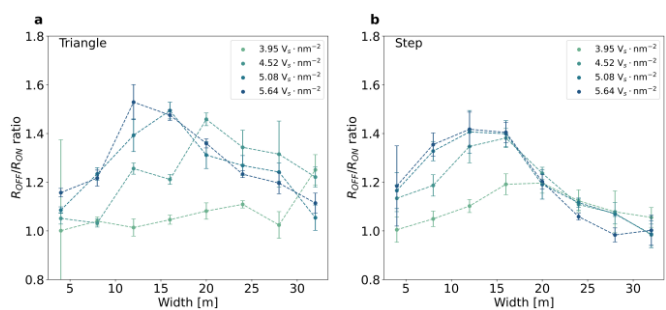

**Figure 6: Device optimization.** The dependence of the resistance ratio on the initial peak density (5.64 $V_s$ nm$^{-2}$, 5.08 $V_s$ nm$^{-2}$, 4.52 $V_s$ nm$^{-2}$ and 3.95 $V_s$ nm$^{-2}$) and device width for two density profiles: (a) a triangle and (b) step distributions. The devices are stressed under consecutive voltage ramps with a rate of 2.1 V s$^{-1}$ between 25.2 to -25.2 V. The resistance ratio is averaged over 15 cycles and the error is the standard deviation of the cycle-to-cycle variability.

View Article Online
DOI: 10.1039/D2NR06425A

conduction mechanism during the ON state is a distributed one, which explains the drop of resistance when the device size in the $y$-axis is increased. We have also assessed different distributions of defect density in the channel to find some routes for device optimization. Comparing the skewed Gaussian distribution, the step function and the triangle distribution, we can conclude the triangle shape enables larger resistance ratios for narrower fissure regions. Besides, the fissure width of these distributions also affects the resistance ratio.

## Author Contributions

J. J. fabricated the devices and collected the experimental data. S. A. developed the simulator and conducted the simulation. S. A. and H. Z. analysed the data and wrote the manuscript. H.Z. conceived the study and supervised the project. All authors have given approval to the final version of the manuscript.

## Conflicts of interest

There are no conflicts to declare.

## Acknowledgements

The authors gratefully acknowledge financial support by the Science Foundation Ireland under 20/FFP-P/8727.